\documentstyle[12pt]{article}
\title{Baryogenesis from a Primordial Lepton Asymmetry}
\author{S.~A.~Abel$^{1}$ and K.~E.~C.~Benson$^{2}$ \vspace{0.3cm}
\\$^{1}$Rutherford Appleton Laboratory
\\Chilton, Didcot
\\Oxon OX11 0QX
\\England
\\ \vspace{0.3cm}
\\$^{2}$Theoretical Physics Department
\\Keble Road
\\Oxford OX1 3NP
\\England}
\newcommand{\beq}{\begin{equation}}
\newcommand{\eeq}{\end{equation}}
\newcommand{\bea}{\begin{eqnarray}}
\newcommand{\eea}{\end{eqnarray}}
\newcommand{\lap}{\begin{array}{c} < \\[-2.1ex] \sim \end{array}}
\newcommand{\gap}{\begin{array}{c} > \\[-2.1ex] \sim \end{array}}
\newcommand{\sub}[1]{_{\mbox{\scriptsize #1}}}
\renewcommand{\sup}[1]{^{\mbox{\scriptsize #1}}}

\begin{document}

\maketitle

\title{Abstract}

\begin{abstract}
\noindent We examine the generation of baryon asymmetry at the weak scale from
a primordial lepton asymmetry. If the electroweak phase transition  is first
order, partial reflection of tau leptons off the bubble walls  and the
resulting hypercharge transport into the symmetric phase, gives  rise to a
baryon number consistent with observation.
\end{abstract}
\pagebreak

\section*{Introduction}

Electroweak theory contains the means for anomalous violation of fermion number
via non-perturbative effects, in the form of instantons at zero temperature
\cite{thooft} and sphalerons at high temperature \cite{manton}.  The vacuum
structure of electroweak theory admits both quantum and thermal tunneling
between inequivalent vacua by such configurations.    Because of this, any
theory in which $B-L=0$ is in danger of having its baryon number erased by
sphalerons.

One way around this problem is to suppose that    the baryon  asymmetry was
made at the electroweak scale \cite{turok-zadroz,nkc}. If this is the case the
sphaleron transitions must have fallen out of  equilibrium whilst there was
still some chemical potential for non-zero  baryon number. If the CP violation
is provided by the higgs sector, this implies that the phase transition was at
least mildly  first order \cite{dine,first}.

The alternative possibility which will concern us here, is that the  asymmetry
was encoded in a conserved global quantum number such as  $B-L$ number
\cite{fukugita}.   This was generated  (or set by hand) primordially,
and after possible processing at the  electroweak phase transition, was
converted into the presently observed  baryon number. Recent work
\cite{kuzmin,dreiner} has suggested   that the three  \[ L_i - B/3  \]  numbers
may play this role ($i$ is a generation index). These could have been produced
by generational (and CP violating)  differences in the out of equilibrium
decays of lepto-quarks \cite{kuzmin,luty}. The baryogenesis could have occurred
{\em  after} the   electro-weak phase transition as a results of mass effects.
For this to be possible it is necessary that sphalerons remained in
equilibrium for some time implying a second--order transition.  Campbell,
Davidson, Ellis, and Olive \cite{campbell} have taken this idea further by
demonstrating how to avoid the $m^2 /T^2$ suppression through lepton-violating
interactions.  Provided that some, but not all, lepton flavours are violated by
$\Delta L \neq 0$ interactions in equilibrium, $B$ may be regenerated without
the $m^2 /T^2$ lepton mass effects.  Instead $B$ depends only on the initial
asymmetry of the non-equilibrating modes.  For example, if lepton-violating
interactions equilibrate $L_1 - B/3$ and $L_2 - B/3$ to zero, then the non-zero
$L_3 - B/3$ biases baryon production via sphalerons: \[ B \propto B-L = B/3 -
L_3 \neq 0 \; . \]

The above example is interesting for two reasons. Firstly it can operate even
if $B-L=0$. Secondly there is the novelty that two of the Sakharov  conditions
(C/CP violation and thermal non-equilibrium) are important at  early times,
whilst the third (baryon number violation) is important  at `late' times (where
`late' means at the weak scale;  after $\approx 10^{-10}$ seconds). In the
light of this, we shall discuss an alternative way in which these quantum
numbers can furnish a net baryon number.

Again we rely on some unspecified  mechanism to generate a non-zero $L_3 - B/3$
number. The values of the  other two numbers are unimportant, and it could be
that $B-L=0$.   The phase transition is assumed to be robustly first order, as
argued for  in the case of the Standard Model in refs.\cite{first,farrar}, and
as  demonstrated in extensions of it in ref.\cite{pietroni}. Once the
temperature drops below  the critical temperature, bubbles of true vacuum begin
to nucleate and  expand into the symmetric phase. As the walls advance through
the plasma,  a fraction (proportional to their physical mass) of the
$\tau$-leptons are  reflected back into the symmetric phase. Since the
$L_3-B/3$ number is non-zero, more taus than anti-taus are reflected, and a
small hypercharge builds up  on the outside of the bubbles with an opposing
hypercharge on the  inside. On the outside sphaleron transitions convert some
of the lepton  number into baryons, whilst on the inside sphaleron transitions
are  suppressed by a large exponential factor, resulting in a net  number of
baryons being produced.

Since the reflection is only suppressed  by the factor $M_\tau/T$ one might
expect this mechanism to be much more  efficient than the equilibrium process
described in refs.\cite{kuzmin,dreiner}. We shall see that in fact the two
mechanisms give remarkably similar baryon asymmetries. This is due to an
additional  suppression due to the low rate of conversion of lepton into
baryon number compared to the typical wall velocity.    The resulting baryon
number is proportional to the initial $\tau$-lepton asymmetry $\rho_\tau$. We
stress that we do not make any assumption about  enhanced (or maximal) CP
violation in the dynamics of the phase transition (which plays no role in our
mechanism~\footnote{CP violation is of course  evident in the initial condition
of a primordial L-asymmetry; and CP  violation in the particle dynamics would
have been required for its  generation at the GUT scale, as required by the
Sakharov conditions.}).

At high temperature or density, the fundamental excitations do not coincide
with the elementary particles. As a result, perturbation theory in terms of the
bare fields does not properly describe effects essentially due to the ambient
plasma.  In order to proceed in a valid perturbative calculation, we first
determine the quasiparticle modes that exist in a relativistic plasma at high
temperature.  Using these modes, we compute the reflection coefficients and
then the lepton flux reflected off the expanding bubble wall.  Because the
reflected flux $\propto M_l /T$, where $M_l$ is the lepton flavour mass and
$T\sim 100$ GeV is the critical temperature of the phase transition, the
$\tau$-lepton flux will predominate.  The thermal scattering length of the
leptonic quasiparticles will then be computed, to give an indication of how
long the reflected leptons remain in the unbroken phase before being absorbed
by the advancing front of broken phase.  During its time in the unbroken phase,
the $\tau$-lepton flux biases anomalous baryon violation, which, it is assumed,
are in equilibrium outside the bubble (the region of $\phi =0$). The sphaleron
processes thus produce baryons, which are swallowed into the broken phase and
survive to the present.  We then estimate the generated baryon number and find
that $\rho_B \sim 10^{-10}$ can be accommodated in this scenario, provided that
the primordial $\tau$-asymmetry is $\rho_\tau /s \sim 5\cdot 10^{-5}$, for a
wall velocity $u\approx 0.1$. Scaling by the lepton number density, \[
\frac{n_l - n_{\bar{l}}}{n_l + n_{\bar{l}}} \sim 0.005 \; . \] Our primary
sources are the papers by Nelson, Kaplan, and Cohen \cite{nkc} and by Farrar
and Shaposhnikov \cite{farrar}, denoted NKC and FS respectively.  The starting
point of our analysis is the rate equation describing the approach of baryon
number to equilibrium  \cite{sakita};
\beq
\dot{\rho}_B = - \frac{ \Gamma_B}{T} \; \frac{\partial F} {\partial B} \; .
\label{eq:b-rate}
\eeq
The partial derivative of the free energy is taken with all conserved quantum
numbers held fixed, and as we will compute later, is simply proportional to the
hypercharge;
\beq
\frac{\partial F} {\partial B} = \xi \frac{\rho_Y}{T^2} \; ,
\eeq
where the parameter $\xi$ depends on the model under consideration and will be
determined explicitly later. Integrating the rate equation ahead of the
advancing bubble wall,
\bea
\rho_B &=& - \frac{ \Gamma_B}{T} \int \!\!dt\; \frac{\partial F}
             {\partial B} = - \frac{ \xi \Gamma_B}{T^3}
             \int\limits_{-\infty}^{z/u} \!\!dt\; \rho_Y (z-ut) \nonumber\\
       &=& - \frac{ \xi \Gamma_B f_Y \tau_T} {u T^3} \; ,
\eea
where $f_Y$ is the reflected hypercharge flux and $\tau_T$ the thermal
transport time ($\sim$ thermal scattering length).  Scaling by the entropy
density and recalling the rate for anomalous baryon violation in the symmetric
phase (for three generations)  \cite{sphal-rate,ambjorn},
\beq
\Gamma_B = 3 \kappa \alpha\sub{W}^4 T^4 \; , \nonumber
\eeq
we arrive at our final expression for the observed baryon number,
\beq
\frac{\rho_B}{s} = - \frac{3 \alpha\sub{W}^4 \kappa \xi} {u}
                     \left( \frac{f_Y \tau_T T}{s} \right) \; .
\label{eq:b-final}
\eeq
We consider the ranges $0.1<\kappa <1$ \cite{ambjorn} and $0.1<u<1$ \cite{liu},
and are left to compute $f_Y$, $\xi$, and $\tau_T$, which we do in the
following sections. For the remainder of this discussion, we shall assume  for
concreteness that $B-L=0$, although this is not necessarily the case as stated
previously.

\section*{Reflected Hypercharge flux ($f_Y$)}

Before we can calculate $f_Y$ we need to determine the leptonic modes in  the
thermal plasma (which is characterised by the four-velocity $u^\alpha$)  and
obtain the quasiparticle solutions for left and right chiralities,  which have
different interactions and hence develop different thermal masses. The first
part of this section section summarises basic results;  more complete
treatments may be found in Weldon and Lebedev \cite{weldon}.

The lepton thermal self-energy at one loop is given by
\bea \Sigma (P) &=& i A \int \!\!\frac{d^4k} {(2\pi)^4}\; D_{\mu\nu}
                    (k) \gamma^\mu S(k+P) \gamma^\nu + i B \int \!\!\frac{d^4k}
                    {(2\pi)^4}\; D(k) S(k+P) \\
           &\equiv& -a(\omega, p) \!\not\!P - b(\omega, p) \!\not\!u \; ,
\label{eq:self-en}
\eea
where $a$ and $b$ are functions of the Lorentz invariants
\bea
\omega &\equiv& P \cdot u \\
     p &\equiv& \sqrt{(P \cdot u)^2 - P^2} \; ,
\eea
such that $\omega^2 - p^2 = P^2$.  The functions $a$ and $b$ are computed in
\cite{weldon}:
\bea
a(\omega, p) &=& \frac{\Omega^2} {p^2} \left[ 1 - \frac{\omega}{2p}
                 \log \left( \frac{\omega +p} {\omega -p} \right) \right] \\
b(\omega, p) &=& \frac{\Omega^2} {p} \left[ -\frac{\omega}{p} +
                 \frac{1}{2} \left( \frac{\omega^2}{p^2} -1 \right) \log
                 \left(\frac{\omega +p} {\omega -p} \right) \right] \\
\Omega^2     &=& (A + \frac{1}{2} B) \frac{T^2}{8} \; .
\eea
The constants $A$ and $B$ depend on the lepton chirality and the model under
consideration.  For the minimal Standard Model and the two-doublet model, one
can easily see that
\beq
\begin{array}{ll}
A_L = \frac{3}{4} g^2 + \frac{1}{4} g'^2 \hspace{2em} &;\hspace{2em}
B_L = g_\tau^2                           \\
A_R = g'^2                               \hspace{2em} &;\hspace{2em}
B_R = 2 g_\tau^2
\end{array} \; ,
\eeq
where the Yukawa contributions are identical for both models, since in either
case, the lepton couples to only one scalar doublet. This yields the following
thermal lepton masses:
\bea
\Omega^2_L &=& \frac{1}{8v^2} \left( 2M_W^2 + M_Z^2 + M_\tau^2 \right)
               T^2 \approx (0.209\, T)^2 \\
\Omega^2_R &=& \frac{1}{4v^2} \left( 2\tan^2 \!\theta\sub{W}\, M_W^2 +
               M^2_\tau \right) T^2  \approx (0.127\, T)^2 \; .
\eea
Note that $\Omega_L > \Omega_R$; this is true for all leptons $l$, provided
that $M_l < 116$ GeV.  Also, this result is gauge invariant. The self-energy
(\ref{eq:self-en}) leads to the following fermion propagator:
\beq
S(P) = \left[ (1+a) \!\not\!p + b \!\not\!u \right]^{-1}
     = \left[ (1+a) \!\not\!p + b \!\not\!u \right] / Z \; ,
\eeq
where the denominator is
\beq
Z(\omega, p) = (1+a)^2 P^2 + 2(1+a)b\, P \cdot u + b^2
             = \left[\omega (1+a) + b \right]^2 - \left[ p(1+a) \right]^2 \; .
\eeq
The poles of the propagator occur at the zeros of $Z(\omega,p)$,
\beq
\omega (1+a) + b = \pm p (1+a) \; ,
\eeq
which includes both positive- and negative-energy solutions (note that $(1+a)$
is even, and $b$ odd, under $\omega \rightarrow -\omega$); given a
positive-energy solution $\omega(p)$, the corresponding negative-energy
solution is $-\omega(p)$.  So it suffices to find only the positive-energy
solutions. The dispersion relation is given by
\bea
\omega \mp p &=& -a(\omega,p) (\omega \mp p) - b(\omega,p) \nonumber\\
&=& \frac{\Omega^2}{p} \left[ \pm 1 + \frac{1}{2} \left( 1 \mp
\frac{\omega}{p} \right) \log \left( \frac{\omega+p}{\omega-p} \right)
\right] \; .
\eea
Each chirality has two distinct modes, shown in Figure~1,  which we label
normal ($\omega_+ (p)$) and abnormal ($\omega_- (p)$). The abnormal mode is
actually unstable for $p \gap \Omega$ \cite{weldon}.   We plot the dispersion
relations in Figure~2 for left and right chiralities of the $\tau$-lepton. The
dispersion relations may be approximated for small and large $p$ as
\bea
\frac{\omega_+ (p)}{\Omega} &\approx& \left\{ \begin{array}{ll} 1 +
\frac{1}{3} \frac{p}{\Omega} + \frac{1}{3} \left( \frac{p}{\Omega}
\right)^2 &: p \lap \Omega \\ \sqrt{2 + \left( \frac{p}{\Omega}
\right)^2} &: p \gap \Omega \end{array} \right. \\
\frac{\omega_- (p)}{\Omega} &\approx& \left\{ \begin{array}{ll} 1 -
\frac{1}{3} \frac{p}{\Omega} + \frac{1}{3} \left( \frac{p}{\Omega}
\right)^2 &: p \lap \Omega \\ \frac{p}{\Omega} \left[ 1 + 2\exp \left(
-2 \left( \frac{p}{\Omega} \right)^2 - 2 \right) \right] &: p \gap
\Omega \end{array} \right. \; .
\eea
In the broken phase, the Dirac operator is
\beq
\left( \begin{array}{cc} \Sigma^0_L + \Sigma\sup{b}_L & M \\ M^\dagger
& \Sigma^0_R + \Sigma\sup{b}_R \end{array} \right) \; ,
\eeq
where the thermal piece includes a contribution due to mass
corrections in the broken theory:
\beq
\Sigma\sup{b}_{L,R} = \Sigma\sup{s}_{L,R} + \delta
\Sigma_{L,R} \; .
\eeq
We then obtain the dispersion relation from
\beq
\det \left( \begin{array}{cc} \Sigma^0_L + \Sigma\sup{b}_L & M \\
M^\dagger & \Sigma^0_R + \Sigma\sup{b}_R \end{array} \right) = 0
\; .
\eeq
The Lagrangian may be made linear for $p \ll \Omega_{L,R}$,
\bea
{\cal L}\sub{eff} &=& 2i L^\dagger \left( \partial_0 - \frac{1}{3}
\vec{\sigma} \cdot \vec{\partial} + i\Omega_L \right) L + 2i R^\dagger
\left( \partial_0 + \frac{1}{3} \vec{\sigma} \cdot \vec{\partial} +
i\Omega_R \right) R \nonumber\\
&& \mbox{} + L^\dagger M R + R^\dagger M^\dagger L \; ,
\eea
yielding the dispersion relation
\beq
\omega (p) = \frac{\Omega_L + \Omega_R}{2} \pm
\sqrt{ \frac{M^2}{4} + \left( \frac{\Omega_L - \Omega_R}{2} \pm
\frac{|\vec{p}|}{3} \right)^2 } \; ,
\eeq
where the first $\pm$ is for $L,R$ and the second for normal or abnormal,
respectively; the solutions $\omega(p)$ are shown in Figures~3 and 4. Notice
that the left abnormal and right normal lines do not intersect (as they do in
the symmetric phase), but instead are separated by an energy interval $\Delta
\omega = M_\tau$ about the point $\omega_0 = ( \Omega_L + \Omega_R )/2$.  This
interval is the region of total reflection, as we will see below when we
consider the scattering of fermions off the bubble wall.

\bigskip
\noindent
In order to calculate the reflection coefficients, we start from the  full
Dirac equation,
\beq
\left( \begin{array}{cc} \Sigma^0_L + \Sigma\sup{s}_L + \delta
\Sigma_L & {\cal M} \\ {\cal M}^\dagger & \Sigma^0_R + \Sigma\sup{s}_R
+ \delta \Sigma_R \end{array} \right) \left( \begin{array}{c} L \\ R
\end{array} \right) = 0 \; ,
\eeq
where ${\cal M} (x)$ is the position-dependent mass and embodies the details of
the bubble wall profile; in the broken phase $\Sigma\sup{b} = \Sigma\sup{s} +
\delta \Sigma$ and ${\cal M} = M$, while in the symmetric phase $\delta \Sigma
= 0 = {\cal M}$. It is because the left and right chiralities interact
differently  with the bubble wall that there exists the possibility of
seperating C- and CP-odd  reflecting currents (weak hypercharge for instance).

As a first approximation, we set $\Sigma\sup{b} \approx \Sigma\sup{s}$ or
$\delta \Sigma \approx 0$, which is plausible since for leptons $M/T \ll 1$;
$\delta \Sigma$ includes mass corrections in the one-loop graphs via the
propagator $[\!\not\!p + M]^{-1}$, where the momentum integrals are dominated
by the region $p \sim T$. Again we may linearise  the Dirac equation at low
momenta, $p \ll \Omega$:
\beq
\left( \!\!\begin{array}{cc} \omega \left(1 + \tilde{\alpha}_L +
\tilde{\beta}_L \right) + i \vec{\sigma} \cdot \vec{\partial} \left(
1+ \tilde{\alpha}_L \right) & {\cal M} \\ {\cal M}^\dagger & \omega \left(1 +
\tilde{\alpha}_R + \tilde{\beta}_R \right) - i \vec{\sigma} \cdot
\vec{\partial} \left( 1+ \tilde{\alpha}_R \right) \end{array}\!\!
\right) \!\!\left( \!\!\begin{array}{c} L \\ R \end{array}\!\! \right)
= 0 \; .
\eeq
This may be written as
\beq
\left( \begin{array}{cc} \sigma^j & 0 \\ 0 & - \sigma^j \end{array}
\right) \frac{\partial\Psi} {\partial x^j} = i U R \Psi \; ,
\eeq
where
\bea
U &=& \left( \begin{array}{cc} \omega \left( 1 + \tilde{\alpha}_L +
\tilde{\beta}_L \right) & {\cal M} \\ {\cal M}^\dagger & \omega \left( 1+
\tilde{\alpha}_R + \tilde{\beta}_R \right) \end{array} \right) \\
R &=& \left( \begin{array}{cc} \left( 1 + \tilde{\alpha}_L
\right)^{-1} & 0 \\ 0 & \left( 1+ \tilde{\alpha}_R \right)^{-1}
\end{array} \right) \\
\Psi &=& R^{-1} \left( \begin{array}{c} L \\ R \end{array} \right) \\
\tilde{\alpha}_L &=& - \frac{1}{3} \frac{\Omega^2_L} {\omega^2} \left(
1-3u -2u^2 \right) (1-u) \hspace{1.5em} ; \hspace{1.5em}
\tilde{\beta}_L = - \frac{2}{3} \frac{\Omega^2_L}{\omega^2} (1+u)^2
(1-u) \hspace*{2em} \\
\tilde{\alpha}_R &=& - \frac{1}{3} \frac{\Omega^2_R} {\omega^2} \left(
1+3u -2u^2 \right) (1+u) \hspace{1.5em} ; \hspace{1.5em}
\tilde{\beta}_R = - \frac{2}{3} \frac{\Omega^2_R}{\omega^2} (1-u)^2
(1+u) \hspace*{2em} \; .
\eea
A Lorentz transformation from the fluid frame to the wall frame (with velocity
$u$) has been performed in the small $u$ limit; the $u$-dependence of
$\tilde{\alpha}$ and $\tilde{\beta}$ reflect the spinor transformation.
Because we are working in the wall frame, the energy and momentum parallel to
the wall are conserved: $i \frac{d}{dt} = \omega$, $i \frac{\partial}{\partial
x_{||}} = p_{\|}$. The reflection coefficient depends strongly only on $p_\bot$
(taken to be $p_z$), and so we set $p_\| =0$:
\beq
\left( \begin{array}{cc} \sigma^3 & 0 \\ 0 & - \sigma^3 \end{array}
\right) \frac{\partial\Psi} {\partial z} = i U R \Psi \; ,
\eeq
which decomposes into
\bea
\frac{\partial}{\partial z} \left( \begin{array}{c} \psi_1 \\ \psi_3
\end{array} \right) &=& i \left( \begin{array}{cc} \omega \left(
\frac{1 + \tilde{\alpha}_L + \tilde{\beta}_L} {1 + \tilde{\alpha}_L}
\right) & {\cal M} \left( \frac{1} {1+\tilde{\alpha}_R} \right) \\ -
{\cal M}^\dagger \left( \frac{1}{1+ \tilde{\alpha}_L} \right) & - \omega
\left( \frac{ 1 + \tilde{\alpha}_R + \tilde{\beta}_R } {1 +
\tilde{\alpha}_R} \right) \end{array} \right) \left( \begin{array}{c}
\psi_1 \\ \psi_3 \end{array} \right) \\
\frac{\partial}{\partial z} \left( \begin{array}{c} \psi_4 \\ \psi_2
\end{array} \right) &=& i \left( \begin{array}{cc} \omega \left(
\frac{1 + \tilde{\alpha}_R + \tilde{\beta}_R} {1 + \tilde{\alpha}_R}
\right) & {\cal M}^\dagger \left( \frac{1} {1+ \tilde{\alpha}_L} \right) \\ -
{\cal M} \left( \frac{1}{1+ \tilde{\alpha}_R} \right) & - \omega \left(
\frac{1 + \tilde{\alpha}_L + \tilde{\beta}_L } {1 + \tilde{\alpha}_L}
\right) \end{array} \right) \left( \begin{array}{c} \psi_4 \\ \psi_2
\end{array} \right) \label{eq:dirac_24} \; .
\eea
This describes reflection and transmission perpendicular to the wall. In
analogy to the method of NKC, we write the solution as a path-ordered integral,
\bea
\left( \begin{array}{c} \psi_1 \\ \psi_3 \end{array} \right) (z,t) &=&
e^{-i \omega t}\; \Omega (z) \left( \begin{array}{c} \psi_1 \\ \psi_3
\end{array} \right)_0 \\
\left( \begin{array}{c} \psi_4 \\ \psi_2 \end{array} \right) (z,t) &=&
e^{-i \omega t}\; \overline{\Omega} (z) \left( \begin{array}{c} \psi_4
\\ \psi_2 \end{array} \right)_0 \; ,
\eea
where
\bea
\Omega (z) &=& {\cal P} \exp i \int\limits_{-z_0}^z dx \left(
\begin{array}{cc} \omega \left( \frac{1 + \tilde{\alpha}_L +
\tilde{\beta}_L} {1 + \tilde{\alpha}_L} \right) & {\cal M} \left( \frac{1}
{1+ \tilde{\alpha}_R} \right) \\ - {\cal M}^\dagger \left( \frac{1} {1+
\tilde{\alpha}_L} \right) & - \omega \left( \frac{1 + \tilde{\alpha}_R
+ \tilde{\beta}_R} {1 + \tilde{\alpha}_R} \right) \end{array} \right)
\\
\overline{\Omega} (z) &=& {\cal P} \exp i \int\limits_{-z_0}^z dx
\left( \begin{array}{cc} \omega \left( \frac{1 + \tilde{\alpha}_R +
\tilde{\beta}_R} {1 + \tilde{\alpha}_R} \right) & {\cal M}^\dagger \left(
\frac{1} {1+ \tilde{\alpha}_L} \right) \\ - {\cal M} \left( \frac{1} {1+
\tilde{\alpha}_R} \right) & - \omega \left( \frac{1 + \tilde{\alpha}_L
+ \tilde{\beta}_L} {1 + \tilde{\alpha}_L} \right) \end{array} \right)
\; .
\eea
We denote the path-ordered exponentials by
\bea
\Omega (z) &=& {\cal P} \exp i \int\limits_{-z_0}^z \!\!dx\; Q (x) \\
\overline{\Omega} (z) &=& {\cal P} \exp i \int\limits_{-z_0}^z
\!\!dx\; \overline{Q} (x) \; .
\eea
Note that $\Omega$ and $\overline{\Omega}$ satisfy the differential
equations $\frac{d \Omega}{dz} = iQ$ and $\frac{d \overline{\Omega}}
{dz} = i \overline{Q}$.  $\Omega$ describes $L \rightarrow R$
reflection and $R \rightarrow R$ transmission, while
$\overline{\Omega}$ describes $R \rightarrow L$ reflection and $L
\rightarrow L$ transmission.
As a first approximation, we take the wall profile to be described by
a step-function:
\beq
{\cal M} = \left\{ \begin{array}{ll} M &: z > 0 \\ 0 &: z < 0
\end{array} \right. \; .
\eeq
Then in the symmetric phase,
\beq
\Omega (z<0) = \left( \!\!\begin{array}{cc} \exp i\omega \left(
\frac{1 + \tilde{\alpha}_L + \tilde{\beta}_L} {1 + \tilde{\alpha}_L}
\right) z \!&\! 0 \\ 0 \!&\! \exp -i\omega \left( \frac{1 +
\tilde{\alpha}_R + \tilde{\beta}_R} {1 + \tilde{\alpha}_R} \right) z
\end{array}\!\! \right) \equiv \left( \!\!\begin{array}{cc} \exp i
p\sup{s}_L z \!&\! 0 \\
0 \!&\! \exp -i p\sup{s}_R z \end{array}\!\! \right) \; ,
\eeq
while in the broken phase, $\Omega$ may be diagonalised as
\beq
\Omega (z>0) = D^{-1} \left( \begin{array}{cc} \exp i p\sup{b}_L z & 0 \\ 0
& \exp -i p\sup{b}_R z \end{array} \right) D \; ,
\eeq
where the momenta in the broken phase are
\beq
\pm p\sup{b}_{L,R} = \frac{p\sup{s}_L - p\sup{s}_R} {2} \pm \sqrt{
\left( \frac{p\sup{s}_L + p\sup{s}_R}{2} \right)^2 - \frac{M^2} {(1+
\tilde{\alpha}_L) (1+ \tilde{\alpha}_R) } } \; .
\label{eq:mom-broken}
\eeq
Then
\beq
D \left( \begin{array}{c} \psi_1 \\ \psi_3 \end{array} \right) (z > 0)
= \left( \begin{array}{cc} \exp ip\sup{b}_L z & 0 \\ 0 & \exp
-ip\sup{b}_R z \end{array} \right) D \left( \begin{array}{c} \psi_1 \\
\psi_3 \end{array} \right) (0) \; ,
\eeq
with the diagonalisation matrix given by
\beq
D = \left( \begin{array}{cc} \frac{1}{Y} (\overline{\omega} + \sqrt{B}) &
\frac{1}{Y} \left( \frac{M}{1 + \tilde{\alpha}_R} \right) \\
\frac{1}{X} \left( \frac{M}{1 + \tilde{\alpha}_L} \right) &
\frac{1}{X} \left( \overline{\omega} + \sqrt{B} \right) \end{array}
\right) \; ,
\eeq
where
\bea
\overline{\omega} &=& \frac{p\sup{s}_L + p\sup{s}_R} {2} \\
B &=& \overline{\omega}^2 - \frac{M^2} { (1+ \tilde{\alpha}_L) (1+
\tilde{\alpha}_R) } \; .
\eea
Here $X$ and $Y$ are just normalisation constants.  The reflection coefficient
comes from the condition
\beq
\left( \begin{array}{c} T \\ 0 \end{array} \right) = D \left(
\begin{array}{c} 1 \\ R \end{array} \right) \; ,
\eeq
yielding
\beq
{\cal R}_{L \rightarrow R} = \left| R \right|^2_{L \rightarrow R} =
\left| \frac{1 + \tilde{\alpha}_R} {1+ \tilde{\alpha}_L} \right|
\left| \frac{\overline{\omega} - \sqrt{B}} {\overline{\omega} + \sqrt{B}}
\right| \; .
\eeq
Similarly, the right-to-left reflection coefficient may be found from the Dirac
equation (\ref{eq:dirac_24}) for $\psi_2$ and $\psi_4$, with the result
\beq
{\cal R}_{R \rightarrow L} = \left| \frac{1 + \tilde{\alpha}_L} {1+
\tilde{\alpha}_R} \right| \left| \frac{\overline{\omega} - \sqrt{B}}
{\overline{\omega} + \sqrt{B}} \right| \; .
\eeq
Note that ${\cal R}_{L \rightarrow R} \approx {\cal R}_{R \rightarrow L}
\approx 1$ if $\sqrt{B}$ is imaginary; that is, if
\beq
\overline{\omega}^2 < \frac{M^2}{ (1+ \tilde{\alpha}_L) (1+
\tilde{\alpha}_R) } \; ,
\eeq
we get {\it total reflection}.  In physical terms, an imaginary contribution to
the momentum (\ref{eq:mom-broken}) implies an evanescent (decaying exponential)
transmission amplitude in the broken phase.

Although the step-function wall profile may produce spurious effects due to its
discontinuity, Farrar and Shaposhnikov considered the smooth profile
\beq
{\cal M}^2 = \frac{M^2}{1+ e^{-az}} \; ,
\eeq
and found the reflection coefficients
\bea
{\cal R}_{L \rightarrow R} &=& \left| \frac{1 + \tilde{\alpha}_R} {1+
\tilde{\alpha}_L} \right| \left| \frac{ \sinh \frac{\pi}{a}
(\overline{\omega} - \sqrt{B}) } { \sinh \frac{\pi}{a}
(\overline{\omega} + \sqrt{B}) } \right| \nonumber\\
{\cal R}_{R \rightarrow L} &=& \left| \frac{1 + \tilde{\alpha}_L} {1+
\tilde{\alpha}_R} \right| \left| \frac{ \sinh \frac{\pi}{a}
(\overline{\omega} - \sqrt{B}) } { \sinh \frac{\pi}{a}
(\overline{\omega} + \sqrt{B}) } \right| \; .
\eea
Notice that total reflection again appears for $\overline{\omega}^2 < M^2 / (1
+ \tilde{\alpha}_L) (1 + \tilde{\alpha}_R)$.  We will take this as a general
condition for total reflection (like the condition $\omega < M$ for the
zero-temperature case).  The region of total reflection is shown in Figure~5.
As expected by comparison to the dispersion relations in the broken phase
(Figure~4), total reflection occurs in the energy interval of width $\Delta
\omega \sim M$ about $\omega_0 \sim (\Omega_L + \Omega_R)/2$; the deviation
results from the dependence of the reflection coefficients on the wall velocity
$u$.

\bigskip
\noindent
Using this result we are now able to compute the hypercharge flux  reflected
off the bubble wall.  We neglect quarks, since we assume  that the primordial
baryon asymmetry vanishes and that CP violation  in the CKM matrix is
inadequate to produce the observed baryon number.   We need to calculate
\bea
f_Y &=& - \frac{1}{2} f_{\tau_L} + \frac{1}{2} f_{\bar{\tau}_R} -
f_{\tau_R} + f_{\bar{\tau}_L} - \frac{1}{2} f_{\nu_\tau} + \frac{1}{2}
f_{\bar{\nu}_\tau} \nonumber\\
&& \mbox{} + (\mbox{$\mu$-contribution}) + (\mbox{$e$-contribution})
\; .
\eea
Because $M_\tau \gg M_\mu, M_e$, the interactions of the muon and electron
families with the bubble wall are negligible compared to those of the tau
family, and hence their contributions to the reflected flux may be ignored.  In
the following, $f_{\tau_L}$ denotes the $\tau_L$ particle flux in the fluid
frame, while $f_{L}\sup{s,b}$ denotes the $\tau_L$ number density distributions
in the wall frame, for the symmetric and broken phases; and similarly for the
other species. To find $f_j$, the particle flux in the thermal (fluid) frame,
we compute the flux $\gamma f_j$ in the wall frame ($\gamma = 1/ \sqrt{1-u^2}$,
where $u$ is the wall velocity):
\bea
\gamma f_{\tau_L}       &=& \int \!\!\frac{d^3k}{(2\pi)^3}\; \left[
f\sup{s}_R (k_L, k_T)       \cdot{\cal R}_{R\rightarrow L}(\omega)
\right] \\
\gamma f_{\bar{\tau}_R} &=& \int \!\!\frac{d^3k}{(2\pi)^3}\; \left[
f\sup{s}_{\bar{L}}(k_L, k_T)\cdot{\cal R}_{L\rightarrow R}(\omega)
\right] \; .
\eea
The integrals are taken over particle momenta in the wall frame and by CPT and
Lorentz invariance, ${\cal R}_{\bar{L} \rightarrow \bar{R}} = {\cal R}_{L
\rightarrow R}$.   There is no term for reflection off the symmetric phase
because  the dispersion curves for left-abnormal and right-normal modes
intersect in the symmetric phase; see Figure~2. Also there is no term for
particle transmission from the broken phase  since it is assumed that both the
broken and symmetric phases have zero net hypercharge, so that the transmitted
hypercharge flux  is zero when summed over all particle species.   Thus
reflection yields the only substantial hypercharge flux, and in particular, the
contribution from total reflection dominates. The difference of these integrals
may be  written as
\beq
f_{\tau_L} - f_{\bar{\tau}_R} = \frac{1}{\gamma} \int
\!\!\frac{d^3k} {(2\pi)^3}\; \left[ f\sup{s}_R \left( {\cal R}_{R
\rightarrow L} - {\cal R}_{L \rightarrow R} \right) + \left(
f\sup{s}_R - f\sup{s}_{\bar{L}} \right) {\cal R}_{L \rightarrow R}\right] \;.
\eeq
The first term is much smaller (by a factor of $\sim 10$) than   the second
since the difference in reflection coefficients results from the  difference in
the left and right thermal masses. We shall therefore omit it in the remainder
of this discussion for the sake of clarity.

The above integral is dominated by the region of total reflection, ${\cal R}
\approx 1$, which (as we have seen above) occurs for $B<0$ or
\beq
\overline{\omega}^2 < \frac{M^2}{ (1+ \tilde{\alpha}_L) (1+
\tilde{\alpha}_R) } \; ,
\eeq
and the hypercharge flux is therefore governed by the $\tau$-lepton;
\bea
f_Y &\approx& - \frac{1}{2} \left( f_{\tau_L} -
f_{\bar{\tau}_R} \right) - \left( f_{\tau_R} - f_{\bar{\tau}_L}
\right) \nonumber\\
&\approx& - \frac{1}{\gamma} \int \!\!\frac{d^3k} {(2\pi)^3}\;
\left\{ \frac{1}{2} \left( f\sup{s}_R - f\sup{s}_{\bar{L}} \right)
\cdot {\cal R}_{L \rightarrow R} + \left( f\sup{s}_L -
f\sup{s}_{\bar{R}} \right) \cdot {\cal R}_{R \rightarrow L} \right\}
\nonumber\\
&\approx& - \frac{1}{\gamma} \int\limits_{B<0} \!\!\frac{d^3k}
{(2\pi)^3}\; \left\{ \frac{1}{2} \left( f\sup{s}_R -
f\sup{s}_{\bar{L}} \right) + \left( f\sup{s}_L - f\sup{s}_{\bar{R}}
\right) \right\} \; .
\label{eq:tflux}
\eea
The flux distributions are taken in the wall frame;
\beq
f\sup{s}_j = \frac{\partial \omega_j}{\partial k_z} \cdot n_F
\left( \gamma \left[ \omega_j - u P^z_j \right] \mp \mu_j \right) \; ,
\eeq
where the group momentum and fermion particle distribution (in the thermal
frame) are
\bea
P^z_j &=& \omega_j \frac{\partial \omega_j}{\partial k_z} \\
n_F (\omega \mp \mu) &=& \frac{1}{e^{(\omega \mp \mu)/T} + 1} \; .
\eea
Substituting into eq.(\ref{eq:tflux}) gives
\bea
f_Y &\approx& - \frac{1}{\gamma} \int\limits_{B<0}
\!\!\frac{d^3k} {(2\pi)^3}\; \left\{ \frac{1}{2} \frac{\partial
\omega_R} {\partial k_z} \left[ n_F \left( \gamma [\omega_R -u P^z_R]
- \mu_{\tau_R} \right) - n_F \left( \gamma [\omega_R -u P^z_R] +
\mu_{\tau_R} \right) \right] \right. \nonumber\\
&& \left. \mbox{\hspace{6.5em}} + \frac{\partial \omega_L}{\partial
k_z} \left[ n_F \left( \gamma [\omega_L -u P^z_L] - \mu_{\tau_L}
\right) - n_F \left( \gamma [\omega_L -u P_L^z] + \mu_{\tau_L} \right)
\right] \right\} \; . \nonumber\\
\eea
As a conservative estimate, we truncate the region of phase space where $B<0$
to the region where $k_\| \lap \frac{3}{2} (\Omega_L - \Omega_R)$ and $\partial
\omega_{L,R} / \partial k_\| \approx 0$ (see FS for a discussion on this
point). The $k_\|$- and $k_z$-integrals can now be separated, giving the result
\bea
f_Y  &\approx& - \frac{9}{16\pi \gamma} \Delta \omega \left( \Omega_L -
\Omega_R \right)^2 \left\{ \frac{1}{2} \Delta n_F \left( \gamma
\left[ \omega_0 - u P^z_R \right] \mp \mu_{\tau_R} \right) \right.
\nonumber\\
&& \left. \rule{0cm}{3ex} \mbox{\hspace{10.5em}} + \Delta n_F \left(
\gamma \left[ \omega_0 - u P^z_L \right] \mp \mu_{\tau_L} \right)
\right\} \; ,
\label{eq:y-flux_trunc}
\eea
where the region of total reflection is centered about $\omega_0$ with spread
$\Delta \omega$.  To lowest order in the chemical potentials
$\mu_{\tau_L,\tau_R} /T$,
\bea
\Delta n_F \left( \gamma \left[ \omega_0 - u P^z_j \right] \mp \mu_j
\right) &=& n_F \left( \gamma \left[ \omega_0 - u P^z_j \right] - \mu_j
\right) - n_F \left( \gamma \left[ \omega_0 - u P^z_j \right] + \mu_j
\right) \nonumber\\
&\approx& \frac{\mu_j/T}{ 1 + \cosh \left( \frac{\gamma (\omega_0 -u
P^z_j)}{T} \right) } \; ,
\eea
and in the limit of small $u$ and $k_\|$,
\beq
P^z \approx \frac{\omega_0}{3} (1 + 2u/3) \; ,
\eeq
hence
\beq
\frac{1}{2} \Delta n_F \left( \gamma \left[ \omega_0 - u P^z_R
\right] \mp \mu_{\tau_R} \right) + \Delta n_F \left( \gamma \left[
\omega_0 - u P^z_L \right] \mp \mu_{\tau_L} \right) \approx
\frac{1}{2} \cdot \frac{(\mu_{\tau_R} + 2 \mu_{\tau_L})/T} { 1 + \cosh
\left( \frac{\omega_0}{T} (1 -u/3) \right) } \; .
\eeq
Noticing that $\rho_\tau = (\mu_{\tau_R} + 2 \mu_{\tau_L}) T^2 /6$ (refer to
the following section), we find that $f_Y$ is simply proportional to the
primordial $\tau$-lepton asymmetry.  Our final result for the reflected
hypercharge flux is then
\beq
f_Y \approx - \frac{27}{16\pi} \cdot \frac{(\Omega_L -
\Omega_R)^2 \Delta \omega}{T^3} \cdot \frac{\rho_\tau} { 1 + \cosh \left(
\frac{\omega_0}{T} (1 - u/3) \right) } \; .
\label{eq:y-flux}
\eeq
We believe this expression to be a conservative lower bound, since we have
truncated the flux integral (\ref{eq:y-flux_trunc}) in the region of total
reflection and we have neglected the contribution of $k_\| \sim \omega_0$.

\section*{Partial Derivative of the Free Energy ($\partial F/ \partial B$)}

We now compute $\partial F/ \partial B$, the partial derivative of the free
energy with all conserved quantum numbers held fixed.  We cannot simply adopt
the result of NKC, since their analysis implies not only $B=L=0$, but also $B_j
= L_j =0$ for the individual generations. Since we are interested in the case
of generational differences in lepton asymmetries, we redo the analysis.

$\partial F / \partial B$ depends on the interactions and species in
equilibrium.  We assume that on the timescale $\tau_T$,  $SU(3)_C \otimes
SU(2)_L \otimes U(1)_Y$ interactions are in equilibrium (including quark mixing
and light-fermion Yukawa interactions, which are assumed to be out of
equilibrium in the NKC analysis), and that only the anomalous $B+L$-violation
is not in equilibrium. The (approximately) conserved quantum numbers and their
associated chemical potentials are then
\bea
B/3 - L_j &\leftrightarrow& \mu_j \nonumber\\
B &\leftrightarrow& \mu_B \nonumber\\
Y/2 &\leftrightarrow& \mu_Y \nonumber\\
T_3 &\leftrightarrow& \mu_T \; . \nonumber
\eea
This implies that on the timescale of interest, $Q$, $L_j$, and $B\pm L$ are
also conserved.  In contrast NKC had $B_1=B_2$ and $B_3$ separately conserved,
and $L_j = L/3$; when the constraints $B=0=L$ and $B_1=B_2=0$ were imposed, one
obtained $B_j = 0 = L_j$.  In our case we have $L_j$ conserved and $B_j = B/3$,
and the constraints we impose are $B=0=L$, so that $B_j=0$ although $L_j \neq
0$ is allowed. The net particle number densities are
\bea
\rho_{t_L} &=& 3 \left[ \frac{1}{9} \left( \mu_1 + \mu_2 + \mu_3
\right) + \frac{1}{3} \mu_B + \frac{1}{6} \mu_Y + \frac{1}{2} \mu_T
\right] \frac{T^2}{6} \nonumber\\
\rho_{b_L} &=& 3 \left[ \frac{1}{9} \left( \mu_1 + \mu_2 + \mu_3
\right) + \frac{1}{3} \mu_B + \frac{1}{6} \mu_Y - \frac{1}{2} \mu_T
\right] \frac{T^2}{6} \nonumber\\
\rho_{t_R} &=& 3 \left[ \frac{1}{9} \left( \mu_1 + \mu_2 + \mu_3
\right) + \frac{1}{3} \mu_B + \frac{2}{3} \mu_Y \right] \frac{T^2}{6}
\nonumber\\
\rho_{b_R} &=& 3 \left[ \frac{1}{9} \left( \mu_1 + \mu_2 + \mu_3
\right) + \frac{1}{3} \mu_B - \frac{1}{3} \mu_Y \right] \frac{T^2}{6}
\nonumber\\
\rho_{c_L} &=& \rho_{u_L} = \rho_{t_L} \hspace{3em} ; \hspace{3em}
\rho_{s_L} = \rho_{d_L} = \rho_{b_L} \nonumber\\
\rho_{c_R} &=& \rho_{u_R} = \rho_{t_R}  \hspace{3em} ; \hspace{3em}
\rho_{s_R} = \rho_{d_R} = \rho_{b_R} \nonumber\\
\rho_{e_L} &=& \left( - \mu_1 - \frac{1}{2} \mu_Y - \frac{1}{2} \mu_T
\right) \frac{T^2}{6} \hspace{3em} ; \hspace{3em} \rho_{\nu_e} =
\left( - \mu_1 - \frac{1}{2} \mu_Y + \frac{1}{2} \mu_T \right)
\frac{T^2}{6} \nonumber\\
\rho_{\mu_L} &=& \left( - \mu_2 - \frac{1}{2} \mu_Y - \frac{1}{2}
\mu_T \right) \frac{T^2}{6} \hspace{3em} ; \hspace{3em} \rho_{\nu_\mu}
= \left( - \mu_2 - \frac{1}{2} \mu_Y + \frac{1}{2} \mu_T \right)
\frac{T^2}{6} \nonumber\\
\rho_{\tau_L} &=& \left( - \mu_3 - \frac{1}{2} \mu_Y - \frac{1}{2} \mu_T
\right) \frac{T^2}{6} \hspace{3em} ; \hspace{3em} \rho_{\nu_\tau} =
\left( - \mu_3 - \frac{1}{2} \mu_Y + \frac{1}{2} \mu_T \right)
\frac{T^2}{6} \nonumber\\
\rho_{e_R} &=& \left( - \mu_1 - \mu_Y \right) \frac{T^2}{6}
\hspace{1.5em} ; \hspace{1.5em} \rho_{\mu_R} = \left( - \mu_2 - \mu_Y
\right) \frac{T^2}{6} \hspace{1.5em} ; \hspace{1.5em} \rho_{\tau_R} =
\left( - \mu_3 - \mu_Y \right) \frac{T^2}{6} \nonumber\\
\rho_{\phi^+} &=& n \left( \mu_Y + \mu_T \right) \frac{T^2}{6}
\hspace{3em} ; \hspace{3em} \rho_{\phi^0} = n \left( \mu_Y - \mu_T
\right) \frac{T^2}{6} \nonumber\\
\rho_{W^+} &=& 4 \mu_T \frac{T^2}{6} \nonumber \; .
\eea
In the above, $n$ denotes the number of scalar doublets in equilibrium.  Then
\bea
\frac{\rho_Y}{2} &=&
3 \cdot \frac{1}{6} \left( \rho_{t_L} + \rho_{b_L}
\right) + 3 \cdot \frac{2}{3} \rho_{t_R} - 3 \cdot \frac{1}{3}
\rho_{b_R} \nonumber\\
&& \mbox{} - \frac{1}{2} \left( \rho_{e_L} + \rho_{\mu_L} +
\rho_{\tau_L} + \rho_{\nu_e} + \rho_{\nu_\mu} + \rho_{\nu_\tau}
\right) \nonumber\\
&& \mbox{} - \left( \rho_{e_R} + \rho_{\mu_R} + \rho_{\tau_R}
\right) + \frac{1}{2} \left( \rho_{\phi^+} + \rho_{\phi^0} \right)
\nonumber\\
&=& \left[ \frac{8}{3} \left( \mu_1 + \mu_2 + \mu_3 \right) + 2 \mu_B
+ (10+n) \mu_Y \right] \frac{T^2}{6} \nonumber\\
\rho_{T_3}       &=& (10+n) \mu_T \frac{T^2}{6} \nonumber \\
\rho_{B}         &=&
3 \cdot \frac{1}{3} \left( \rho_{t_L} + \rho_{b_L} + \rho_{t_R}
+ \rho_{b_R} \right) \nonumber\\
&=& \left[ \frac{4}{3} \left( \mu_1 + \mu_2 + \mu_3 \right) + 4 \mu_B
+ 2 \mu_Y \right] \frac{T^2}{6} \nonumber\\
\rho_{L}         &=&
\rho_{e_L} + \rho_{\mu_L} + \rho_{\tau_L} + \rho_{\nu_e}
+ \rho_{\nu_\mu} + \rho_{\nu_\tau} + \rho_{e_R} + \rho_{\mu_R} +
\rho_{\tau_R} \nonumber\\
&=& \left[ - 3 \left( \mu_1 + \mu_2 + \mu_3 \right) - 6 \mu_Y \right]
\frac{T^2}{6} \nonumber\\
\rho_{L_j}       &=&
\left( - 3 \mu_j - 2 \mu_Y \right) \frac{T^2}{6} = \left[ -
3 \mu_j + (\mu_1 + \mu_2 + \mu_3) \right] \frac{T^2}{6} \; .
\eea
Imposing the conditions $B=0=L$, we find
\bea
\mu_Y                  &=&   6 \mu_B \nonumber\\
\mu_1 + \mu_2 + \mu_3  &=& -12 \mu_B \nonumber \; ,
\eea
giving
\beq
\frac{\partial F}{\partial B} = \mu_B = \frac{\rho_Y}{(5+n) T^2}
\equiv \frac{\xi \rho_Y}{T^2} \; .
\label{eq:xi}
\eeq

\section*{Thermal Transport Time ($\tau_T$)}

After rebounding off the bubble wall, the reflected lepton flux travels into
the symmetric phase until the advancing front of broken phase captures it,
during which time the corresponding hypercharge current biases baryon
production via anomalous processes.  We now estimate the thermal transport time
$\tau_T$, defined as the average time that a reflected $\tau$-lepton spends in
the plasma prior to absorption by the bubble of broken phase.  Consider
diffusion away from the wall of a particle with velocity $v$ and mean free path
$l$. Capture occurs when the wall intercepts the randomly walking particle ($N$
here is the number of collisions):
\beq
u \tau_T = \left\langle l \sqrt{N} \right\rangle = \left\langle l
\sqrt{\frac{v \tau_T}{l}} \right\rangle \; ,
\eeq
or
\beq
\tau_T = \left\langle \frac{lv}{u^2} \right\rangle \; .
\eeq
We therefore need to calculate the thermal average
\beq
\langle l v \rangle =
\left\langle\frac{v}{\langle n\sigma\rangle_1}\right\rangle_2 \; ,
\eeq
where the subscripts 1 and 2 refer to thermal averages taken over the
rebounding particles and particles in the plasma, respectively. Here $n$ is the
particle number density and $\sigma$ the thermally averaged cross section.  The
thermal transport time may then be written as
\beq
\tau_{T}^{-1}= \sum_2 \frac{u^2 g_1 g_2}{\langle n
\rangle_1} \int\! \frac{d^3\vec{p}_1 \, d^3\vec{p}_2} {(2\pi)^6}\,
\frac{E_1}{|\vec{p}_1|}\frac{\sigma_{12}
(\vec{p}_1, \vec{p}_2)} {(e^{E_1 /T} +1) (e^{E_2 /T}+1)} \; ,
\eeq
where the sum is taken over all interactions of particle 1 with the heat bath
and $g_j$ counts the spin degrees of freedom. Since the leptons only interact
electroweakly, it is reasonable to  approximate the above expression by
considering only the leading contributions from tree-level scattering on mass
shell. Numerical calculation confirms that other contributions are indeed less
significant. In this case, the transition probabilities sum to
\bea
\sum \left| T \right|^2
      &\approx& \frac{e^2}{\sin^2
                \!\theta\sub{W}} \left( \Omega^2_W - \Omega_\tau^2 \right) \\
      &       & \mbox{} + \frac{e^2}{\sin^2 \!\theta\sub{W}\, \cos^2
                \!\theta\sub{W}} \left( 4\sin^4 \!\theta\sub{W}\, \Omega^2_W
                + 8\sin^4 \!\theta\sub{W}\, \Omega^2_\tau - 2\sin^2
                \!\theta\sub{W}\, \Omega_W^2 \right. \\
      &       & \left. \mbox{} \hspace*{8em} - 4\sin^2 \!\theta\sub{W}\,
                \Omega_\tau^2 + \Omega_W^2 - \Omega_\tau^2 \right) \\
      &       & \mbox{}+4e^2 \left( \Omega^2_\gamma +\Omega_\tau^2\right)\; ,
\eea
where we respectively list the contributions of $W$, $Z$, and $\gamma$
scattering in the plasma (higgs scattering is negligible since it is suppressed
by the small Yukawa couplings).  $\Omega_j$ is the thermal mass of particle
species $j$, and we have set $\Omega_\gamma \approx \Omega_W \approx 0.5 T$ and
$\Omega_{\nu_\tau} \approx \Omega_\tau\approx 0.2 T$. The thermal transport
time is then given by
\beq
\tau_T \equiv \frac{x}{u^2T} \sim \frac{100}{u^2 T} \; .
\label{eq:tau}
\eeq
Note that, computing $\langle v \rangle = \langle n v \rangle/\langle
n\rangle$, we find that $v \approx 1$ for most leptons in the plasma, which
confirms that the reflected lepton current is quickly thermalised.

\section*{Discussion}

Taking the expression (\ref{eq:b-final}) for the final baryon number and
substituting in eqs.(\ref{eq:y-flux}), (\ref{eq:xi}), and (\ref{eq:tau}) for
the reflected hypercharge flux, the partial derivative of the free energy with
respect to baryon number, and the thermal transport time, we obtain
\bea
\frac{\rho_B}{s} \!&\approx&\! \frac{81} {16\pi} \cdot
\frac{\kappa \alpha\sub{W}^4} {(5+n)u^3} \cdot \frac{x \left(
\Omega_L - \Omega_R \right)^2 \Delta \omega} {T^3 \left[ 1 + \cosh
\frac{\omega_0} {T} (1-u/3) \right]} \cdot \left( \frac{\rho_\tau} {s}
\right) \nonumber\\
\!&\approx&\! 2.1 \cdot 10^{-6} \cdot \frac{\kappa x} {(5+n) u^3
\left[ 1+ \cosh \frac{\omega_0}{T} (1- u/3) \right]} \cdot \frac{
\left( \Omega_L - \Omega_R \right)^2 M_\tau }{T^3} \cdot \left(
\frac{\rho_\tau} {s} \right) \nonumber\\
\!&\approx&\! 2.1 \!\cdot\! 10^{-6} \cdot \kappa \!\left(
\frac{6}{5+n} \right)\! \!\left( \frac{0.1}{u} \right)^{\!3}\!
\!\left( \frac{100\!\mbox{ GeV}}{T} \right)\! \!\left( \frac{2}{1+
\cosh \frac{\omega_0} {T} (1-u/3)} \right)\! \!\left( \frac{x}{100}
\right)\! \cdot  \!\left( \frac{\rho_\tau}{s} \right) \, , \nonumber\\
\label{eq:b_final}
\eea
where the various parameters have been scaled by their typical values. We
believe eq.(\ref{eq:b_final}) to be a conservative lower bound on the effect of
this mechanism since we have underestimated the flux integrals.  By requiring
the observed baryon number of $\rho_B \sim 10^{-10}$ to be generated in this
manner, we obtain an estimate for the primordial lepton asymmetry,
\beq
\frac{\rho_\tau}{s} \sim 5\cdot 10^{-5} \; ,
\eeq
for the typical values of parameters ($\kappa=1$, $u=0.1$, $x=100$).
This corresponds to
\beq
\frac{n_l - n_{\bar{l}}} {n_l + n_{\bar{l}}} \sim 0.005 \; .
\eeq
It is of interest to compare this constraint with that obtained from
equilibrium scenarios for lepton-to-baryon conversion.  As a generic example,
we consider the analysis of Kuzmin, Rubakov, and Shaposhnikov \cite{kuzmin}.
Taking the requisite large Higgs mass to be $M_H \sim 100$ GeV, and
consequently the sphaleron freeze-out temperature (approximately the critical
temperature) to be $T_* \sim 150$ GeV, we estimate the generated baryon
asymmetry as
\beq
\frac{\rho_B} {s} \approx -\frac{4}{13\pi^2} \frac{m^2_\tau}
{T_*^2} \cdot \left( \frac{\rho_\tau}{s} \right) \approx -4.4 \cdot
10^{-6} \cdot \left( \frac{\rho_\tau}{s} \right) \; .
\eeq
Hence the observed baryon asymmetry may be accounted for in this scheme for a
primordial lepton asymmetry of $\rho_\tau /s \sim 2 \cdot 10^{-5}$, which is
remarkably similar to the value found above despite the $m_\tau^2 /T^2$
suppression.  This is not surprising since, as we have emphasised, the
assumption that baryon violation is in equilibrium (on the timescale of
interest) is an overly optimistic one.

A remark is in order about  the behaviour of eq.(\ref{eq:b_final}) as
$u\rightarrow 0$.   In contrast to NKC, our result diverges unashamedly  at low
wall velocity as $\sim 1/u^3$.  To see why this is acceptable, consider the
extreme case of a stationary wall. In this case the $\tau$-leptons simply
diffuse away from the wall, finally establishing equilibrium once the
hypercharge profile is constant (and different)  on either side. In this sense
the wall is behaving like a semi-permeable  membrane since it forces different
partial pressures of leptons on  the outside and inside of the bubble.
Intuitively, one expects the partial  pressure of leptons in the symmetric
phase to create baryons via sphaleron  transitions until it is fully depleted.
However in this calculation we have made an approximation  by setting the
chemical potentials to be constant. This means an unlimited supply of lepton
number -- hence the divergence.  Our approximation breaks down once the
wall velocity is slow enough  so that the lepton numbers are depleted
significantly. This condition  can be gauged by the ratio of generated density
of baryons to initial density of leptons.  This is a small number  for $u\gap
0.001$ which gives a sufficient range of values for first order phase
transitions (where $u\gap 0.01$ is more reasonable)\cite{liu}. Clearly, in this
scenario, the slower the wall velocity the more efficient the mechanism.

\bigskip
\noindent
To summarise, we have considered the generation of baryons from a primordial
$\tau$-lepton asymmetry. There are two ways in which baryogenesis may occur
which give remarkably similar results. The first is already well known in the
literature,  and assumes that the thermal plasma maintains equilibrium during
the electro-weak phase transition.  Anomalous processes may convert
generational lepton asymmetries into  baryon number, whose final value depends
on the primordial asymmetry.  The effectiveness of this scenario is determined
by the suppression factor $M^2_l (T_*) /T_*^2$ at sphaleron freeze-out.   We
have analysed a second mechanism, which would operate during a first-order
phase transition. Reflection of $\tau$-leptons off the phase separation
boundary may radiate a net hypercharge flux, which then triggers baryon
production as described by the rate equation (\ref{eq:b-rate}).  The
effectiveness of this charge transport mechanism is determined by the strength
of the lepton Yukawa interactions with the bubble wall, through the factor $M_l
/T$.  This mechanism works most efficiently for slow wall velocities.  Although
a comparison between the two depends on the choice of parameter values, there
is a clear trade-off between opposing tendencies: anomalous baryon violation in
equilibrium generates greater net $B$, but risks suppression by lepton mass
effects. We have made several assumptions in deriving the final baryon number,
which we summarise:
\begin{itemize}
\item the reflected hypercharge flux is dominated by the contribution
      due to total reflection;
\item the linearised (low-momentum) Lagrangian is valid in the region
      of total reflection;
\item the wall velocity is non-relativistic (implying that the hypercharge
      current rapidly thermalises);
\item the flux integrals may be approximated by  $k_\|
      \lap \frac{3}{2} (\Omega_L - \Omega_R)$.
\end{itemize}
The last assumption together with the parameters $\kappa$, $u$ and $x$ are the
greatest unknowns. For typical values,  we have found that the observed baryon
number of $\rho_B \sim 10^{-10}$ may be generated by our mechanism, if the
primordial lepton asymmetry is $\rho_\tau /s \sim {\cal O} (10^{-5})$.

Finally, we note that scalar leptons in supersymmetry,  may play a similar role
to the $\tau$-lepton above. After finding the quasiparticle modes and
dispersion relations by diagonalising the mass matrices in both the symmetric
and broken phases, one may analyse the scattering of sleptons off the bubble
wall in the fashion described above.  We expect that a region of total
reflection of width $\Delta \omega \sim M_\tau$ would again yield an $M_\tau
/T$-suppressed contribution to the reflected hypercharge flux, in the manner
demonstrated.  This is in contrast to the equilibrium case examined in
ref.\cite{dreiner}, where the minimal supersymmetric extension of the Standard
Model produces a much greater enhancement due to the large mass splittings of
right and left sleptons.

\vspace{1cm}
\noindent
{\bf\Large Acknowledgement} \hspace{0.3cm}
We would like to thank Herbi Dreiner, Subir Sarkar and Noel Cottingham for
stimulating discussions.

\pagebreak

\pagebreak

\noindent
{\bf\Large FIGURE CAPTIONS} \hspace{0.3cm}

\begin{itemize}

\item[Figure~1 : ]The dispersion relations for normal and abnormal plasma
modes in the symmetric phase.

\item[Figure~2 : ]The dispersion relations for left and right chiralities of
the $\tau$-lepton in the symmetric phase of the Standard Model and
two-doublet model.

\item[Figure~3 : ]The dispersion relations, at low momenta in the broken
phase, for normal and abnormal modes of left and right chiralities of the
$\tau$-lepton in the Standard Model and two-doublet model.

\item[Figure~4 : ]Close-up of the previous figure, magnifying the separation
between the left abnormal and right normal lines.

\item[Figure~5 : ]The region of total reflection (bounded by the curves
plotted) for the $\tau$-lepton in the Standard Model and two-doublet
model, as a function of the wall velocity; the temperature is taken to
be $T = 100$ GeV.

\end {itemize}

\end{document}